\documentclass[iop]{emulateapj}

\usepackage{amsmath}
\usepackage{amssymb}
\usepackage{amstext}
\usepackage{apjfonts}
\usepackage{graphicx}
\usepackage{epsfig}
\usepackage{color}
\usepackage{natbib}
\newcommand{\avg}[1]{\langle #1 \rangle}

\begin{document}

\shortauthors{Schwab et al.}
\shorttitle{Lensing Tests of Gravity and Cosmology}
\title{Galaxy-Scale Strong Lensing Tests of Gravity and Geometric Cosmology: \\
Constraints and Systematic Limitations\altaffilmark{*}}

\author{Josiah Schwab\altaffilmark{1,2}, Adam S. Bolton\altaffilmark{3,4} and Saul A. Rappaport\altaffilmark{5}}

\altaffiltext{1}{M.I.T. Department of Physics and Kavli Institute for Astrophysics and Space Research, 
70 Vassar Street, Cambridge, MA 02139, USA; jschwab@mit.edu}
\altaffiltext{2}{Department of Physics, University of California, 366 LeConte Hall,
Berkeley, CA 94720-7300, USA; jwschwab@berkeley.edu}
\altaffiltext{3}{Beatrice Watson Parrent Fellow, Institute for Astronomy, University of Hawai`i,
2680 Woodlawn Drive, Honolulu, HI 96822, USA;  bolton@ifa.hawaii.edu}
\altaffiltext{4}{Department of Physics and Astronomy, University of Utah, 
115 S 1400 E, Salt Lake City, UT 84112, USA; bolton@physics.utah.edu }
\altaffiltext{5}{M.I.T. Department of Physics and Kavli Institute for Astrophysics and Space Research,
70 Vassar Street, Room 37-602B, Cambridge, MA 02139, USA; sar@mit.edu}

\altaffiltext{*}{Based on observations made with the
NASA/ESA 
\textsl{Hubble Space Telescope}, obtained
at the Space Telescope Science Institute, which is operated
by AURA, Inc., under NASA contract NAS 5-26555.
These observations are associated with programs No.\ 10174, No.\ 10494,
No.\ 10587, No.\ 10798, and No.\ 10886.}


\begin{abstract}
Galaxy-scale strong gravitational lenses with measured stellar velocity dispersions
allow a test of the weak-field metric on kiloparsec scales and a geometric measurement of
the cosmological distance--redshift relation, provided that the
mass-dynamical structure of the lensing galaxies can be independently constrained
to a sufficient degree.
We combine data on 53 galaxy-scale strong lenses from the Sloan Lens ACS
Survey with a well-motivated fiducial set of lens-galaxy parameters
to find (1) a constraint on the post-Newtonian parameter $\gamma = 1.01 \pm 0.05$,
and (2) a determination of $\Omega_\Lambda = 0.75 \pm 0.17$
under the assumption of a flat universe.
These constraints assume
that the underlying observations and priors are free of systematic error.
We evaluate the sensitivity of these results to systematic uncertainties in (1) total mass-profile
shape, (2) velocity anisotropy, (3) light-profile shape, and (4) 
stellar velocity dispersion.  Based on these sensitivities,
we conclude that while such strong-lens samples can in principle provide an
important tool for testing general relativity and cosmology, they are unlikely to yield precision 
measurements of $\gamma$ and $\Omega_\Lambda$ unless the properties of 
the lensing galaxies are independently constrained with substantially greater accuracy than at present.
\end{abstract}

\section{Introduction}

Einstein's theory of general relativity (GR) has been an extremely successful description of gravity. It has 
passed all current experimental tests, most famously Eddington's measurement of light deflection 
during the solar eclipse of 1919 \citep{Eclipse1919}; observation of the gravitational redshift by 
\citet{Pound60}; the successful operation of the Global Positioning Satellites \citep{Ashby02}; 
measurements of the Shapiro delay \citep{Shapiro64, Bertotti03}; and extensive studies of 
relativistic effects in binary radio pulsar systems, including verification of energy loss via 
gravitational waves as observed in the Hulse--Taylor pulsar \citep{Taylor79}. Tests of gravity at ever 
higher precisions continue to be pursued through techniques such as lunar laser ranging, where 
the Earth-Moon separation is precisely measured as a function of time \citep{LLR04}. The 
parameterized post-Newtonian (PPN) framework \citep{Thorne71} provides a systematic, 
quantitative way in which to formulate and interpret tests of gravity.  The post-Newtonian parameter 
traditionally denoted by $\gamma$---a measure of the amount of spatial curvature per
unit mass---is currently constrained to be $\gamma = 1 + (2.1\pm 2.3) \times 
10^{-5}$ \citep{Bertotti03} on solar-system length scales. Within the post-Newtonian 
parameterization of GR, $\gamma$ has 
no scale dependence, and thus constraints on deviations from $\gamma = 1$ on significantly larger 
length scales can provide further tests of the theory.

GR has had further success in its application to physical cosmology. From 
assumptions of homogeneity and isotropy, the evolution of the universe as a function of time can be 
predicted from the knowledge of the densities of its constituents. While it was initially assumed that 
the density of the universe was dominated by matter (including a significant
component of non-baryonic ``dark matter''), the discovery that the universe is expanding 
at an accelerating rate (as traced by type Ia supernovae) has forced a modification of this idea 
\citep{Riess98, Perlmutter99}. It is now thought that the universe is filled with a ``dark energy'', an 
unknown substance with negative pressure. The composition of the universe determines its 
evolutionary history through the FRW metric and the Friedmann equations, and the evolution 
history of the universe in turn sets the distance-redshift relation. Thus, measurements of the 
conversion from redshift to distance constrain the density of dark energy, which we denote simply 
by $\Omega_\Lambda$ in this work (i.e., assuming an equation of state $P=-\rho c^2$).

This paper presents an analysis of the quantitative constraints that can be placed on GR as a 
theory of gravity and on the density of dark energy in the universe using the large and 
homogeneous  set of strong gravitational lens galaxies observed by the Sloan Lens ACS (SLACS) 
Survey collaboration \citep{SLACSI}.  To date, research using the SLACS sample has been 
primarily focused on constraining the internal structure and dynamics of the SLACS lenses under 
the assumption of GR and a $\Lambda$CDM cosmology \citep[e.g.][]{SLACSII, SLACSIII, SLACSIV, 
Koopmans09}.  However, with reasonable prior assumptions and independent measurements of 
the structure of early-type galaxies such as the SLACS lenses, the problem can be inverted and the 
sample used to constrain the parameters of GR and of the cosmology.  The first avenue was 
explored for an initial sample of 15 SLACS lenses by \citet{BRB06}, who found the post-Newtonian 
parameter $\gamma$ to be $0.98 \pm 0.07$ on kiloparsec scales by adopting priors on early-type 
galaxy structure taken from observations of the local universe.  The second avenue,
originally suggested in passing by \citet{Golse02}, was examined at length
by \citet{Grillo08}, who discussed the use of the SLACS lensing systems as a means to determine 
cosmological parameters, finding that the ``concordance'' values of
$(\Omega_M,\Omega_{\Lambda}) \simeq (0.3,0.7)$ fell within their 99\% confidence 
regions. In both cases, 
the general approach consists of using angular ``Einstein radii'' measured from imaging, in 
combination with spectroscopic redshifts and parameterized galaxy mass models, to predict an 
observable stellar velocity dispersion---which explicitly depends upon the parameter of interest, 
whether post-Newtonian or cosmological---and then comparing this predicted value with the 
velocity dispersion measured from spectroscopy.  Alternatively, these techniques can be
characterized as treating strong-lensing Einstein radii as ``standard angles'' calibrated
by the stellar velocity dispersions of the lensing galaxies. (Also see \citet{SLACSVI} for a related cosmological application.)

In this work we re-examine the constraints that can be set on both $\gamma$ and 
$\Omega_\Lambda$ using strong gravitational lensing by early-type galaxies, given
the availability of the expanded SLACS sample presented by \citet{SLACSV}.  Our
final emphasis will not be
as much on the most probable values for these parameters, but rather on the uncertainties  
due to systematic errors made in estimating the properties of the lensing galaxies.
In Section \ref{sec:slacs}, we 
discuss the SLACS lens sample that is used in our analysis.  The lensing formalism that we utilize 
is reviewed in Section \ref{sec:formalism}.
Our determination of $\gamma$ and $\Omega_\Lambda$ for a fiducial set of 
lensing parameters is presented in Section \ref{sec:constraints},
along with a brief discussion of the statistical analysis that 
was employed.  Section~\ref{sec:systematics}
is devoted to an in-depth evaluation of the effects that systematic errors 
in the lens parameters have on the nominal values for $\gamma$ and $\Omega_\Lambda$.  Finally,
in Section \ref{sec:summcon} we 
summarize our findings and draw some conclusions on how to improve the accuracy of 
strong lensing results in the future.

\section{SLACS Lens Sample}
\label{sec:slacs}

The post-Newtonian strong-lensing constraints published by \citet{BRB06} were based on a
sample of 15 strong galaxy-galaxy gravitational lenses from the SLACS
Survey, which are described in \citet{SLACSI}, \citet{SLACSII}, and \citet{SLACSIII}.
The current work uses a more recent and larger sample of strong lenses from this
same survey, which is described in detail by \citet{SLACSV}.  These systems were all selected
from within the spectroscopic database of the Sloan Digital Sky Survey
(SDSS: \citealt{York_SDSS}) based upon the presence of two significantly different galaxy
redshifts within a single spectrum, obtained with a
3$^{\prime\prime}$-diameter fiber aperture.  As a consequence
of explicit lensing and other selection effects, the SLACS sample consists primarily
of massive, early-type (i.e., elliptical and S0) galaxies lensing much fainter and more
distant emission-line galaxies.  

For the present work, the key observables in each system
are the redshifts of the two components (foreground ``lens'' and background ``source''),
the stellar velocity dispersion of the lens galaxy, a power-law index representing the 
characteristic slope of the luminosity profile, and the angular Einstein radius of 
the strongly lensed image of the more distant galaxy. 
 The first three of these quantities are measured from SDSS spectroscopy,
while the last two are measured from high-resolution follow-up images obtained by the SLACS
survey through the F814W ($I$-band) filter with the Wide Field Channel of the Advanced
Camera for Surveys (ACS) aboard the \textsl{Hubble Space Telescope} (\textsl{HST}).
Note that mass and light 
models fitted to the \textsl{HST} data include a projected axis-ratio parameter, $q$, to capture ellipticity. To
connect these models to the axisymmetric approximation of the analytic Jeans equation-based
framework introduced below, we use the interchange
\begin{equation}
R \leftrightarrow R_q = \sqrt{qx^2 + y^2/q} ~~,
\end{equation}
which conserves the total mass or light within a given isodensity or isobrightness contour
and is consistent with the Einstein-radius and effective-radius measurement conventions
of \citet{SLACSV}.

The 53 lenses
considered in this work are those from \citet{SLACSV} that have single lens galaxies with
early-type morphology, successful quantitative strong lens models (and hence measured
Einstein radii), and sufficiently
high signal-to-noise measurements of lens-galaxy stellar velocity dispersion.  These lens
galaxies have redshifts in the range $z\sim$0.1--0.3, and given these relatively low redshifts
(and hence modest look-back times),
there is unlikely to be significant structural difference between the SLACS
lens galaxies and the local-universe elliptical galaxies that we and \citet{BRB06} use
as a calibration sample.

\section{Lensing Formalism}
\label{sec:formalism}

We start with the weak-field PPN form of the Schwarzschild metric
\begin{equation}
d\tau^2 = dt^2 (1-2M/r) - dr^2 (1-2\gamma M/r) - r^2 d\phi^2
\end{equation}
with $G = c = 1$, where $\gamma$ is a parameter to be constrained, and $\gamma=1$ for GR.
For thin gravitational lenses, the ``lens equation''
which governs the deflection of light is then
\begin{equation}
\vec{\theta}_s = \vec{\theta} - \frac{(1 + \gamma)}{2}\vec{\nabla} \psi(\vec{\theta}) ~~~,
\label{eq:lenseq}
\end{equation}
where $\vec{\theta}_s$ is the angular location of the source,  and $\vec{\theta}$ is the angular location of 
the image.  The scaled, projected Newtonian potential $\psi$ is defined as
\begin{equation}
\psi(\vec{\theta}) = \frac{D_{LS}}{D_L D_S} \frac{2}{c^2} \int d\mathcal{Z} \  \Phi(D_L \vec{\theta}, 
\mathcal{Z})  ~~~,
\label{eq:lenspot}
\end{equation}
where $\Phi$ is the traditional Newtonian potential and $\mathcal{Z}$ is the line-of-sight 
distance between the observer and the source, with $\mathcal{Z} = 0$ at the lens.
Here, $D_S$ is the distance 
to the source, $D_L$ is the distance to the lens, and $D_{LS}$ is the distance between the lens and 
the source; all three are cosmological angular-diameter distances.  

For a lens that is cylindrically symmetric around the line of sight, the Einstein ring radius is found by 
setting $\vec{\theta}_s=0$ to yield
\begin{equation}
\theta_E = \frac{(1 + \gamma)}{2}\, \left(\frac{d \psi (\theta)}{d\theta} \right)_E ~~~.
\end{equation}
The Einstein radius provides the characteristic angle (or length) in gravitational lensing. The 
angular size of the Einstein radius corresponding to a point mass $M$ is given by
\begin{equation}
\theta_E = \sqrt{\frac{1+\gamma}{2}}\left(\frac{4G M}{c^2} \frac{D_{LS}}{D_S D_L} \right)^{1/2} ~~~.
\end{equation}
For more general cylindrically symmetric mass distributions with respect to the line-of-sight, this 
continues to hold with $M = M_E \equiv M(R_E)$, the mass contained within a cylinder 
of radius equal to the Einstein radius. Rearranging terms,
and noting that $R_E = D_L \theta_E$, we find:
\begin{equation}
\frac{G M_E}{R_E} = \frac{2}{(1+\gamma)}\frac{c^2}{4} \frac{D_S}{D_{LS}} \theta_E~~~,
\label{eq:einrad}
\end{equation}
a form which will prove useful. Regardless of the extent of the mass distribution, only the mass 
interior to the Einstein radius has a net effect on the deflection of the light in the
circularly symmetric (or, more generally, homoeoidal \citep{Schramm90}) case.

We will mainly consider a flat cosmology, where $\Omega_k = 0$ and thus $\Omega_M + \Omega_
\Lambda = 1$. This means there is only one free cosmological parameter appearing in the 
distance-redshift relation. This restriction on the dimension of the parameter space yields a sharper 
constraint (see Section \ref{sec:constraints}). The assumption of flatness is not an unreasonable one. 
Currently, independent measurements show $|1 - \Omega_M - \Omega_\Lambda| < 0.02$ \citep
{WMAP5}, and the theoretical prediction of inflation is that $\Omega \approx 1$ \citep{Guth81}.  

In this flat case, the relation among the angular diameter distance between two objects ($D_{A12}
$) and their line-of-sight comoving distances $D_C$ is
\begin{equation}
D_{A12} = \frac{1}{1+z_2}\left(D_{C2} - D_{C1}\right)
\end{equation}
where $z$ is the redshift. Then, the ratio of angular diameter distances $D_{LS}/D_S$ which 
appears in our expression for the Einstein radius is simply
\begin{equation}
\frac{D_{LS}}{D_S} = 1 - \frac{D_{CL}}{D_{CS}}
\label{eq:Dratio}
\end{equation}
Since this is a ratio of two comoving distances, there is no dependence on the Hubble constant. 
Thus, this ratio depends only on the source and lens redshifts, which are measured, and the choice 
of $\Omega_\Lambda$.

One of the simplest models one can write down for an elliptical galaxy is a scale-free model based 
on power-law density profiles for the total mass density, $\rho$, and luminosity density, $\nu$,
\begin{eqnarray}
\label{eq:rhopl}
\rho(r) &=& \rho_0 \left(\frac{r}{r_0}\right)^{-\alpha} \\
\nu(r) &=& \nu_0 \left(\frac{r}{r_0}\right)^{-\delta}
\label{eq:nupl}
\end{eqnarray}
\citep[see, e.g.,][]{Koopmans06}. In general, the three-dimensional
velocity dispersion tensor is not isotropic. The deviation 
can be characterized through the anisotropy parameter $\beta$, defined as
\begin{equation}
\beta(r) = 1 - {\sigma^2_t} / {\sigma^2_r}
\label{eq:beta}
\end{equation}
where $\sigma_r$ is the radial value. Having assumed spherical symmetry, $
\sigma^2_t \equiv \sigma^2_{\theta} = \sigma^2_{\phi}$, and we refer to this component as 
``tangential''.  For the purpose of the current analysis it will suffice to assume that $\beta$ is 
independent of $r$. This is the model used by by \citet{BRB06} and discussed by \citet{Koopmans06}.

We define $r$ to be the spherical radial coordinate from the lens center and $R$ to be the 
cylindrical radius, i.e., perpendicular to the line of sight (defined to be along the 
$\mathcal{Z}$-axis). So by definition $r^2 = R^2 + \mathcal{Z}^2$. 

A key ingredient in all the lensing measurements is the {\em observed} velocity dispersion, 
which is a projected, luminosity weighted average of the radially-dependent velocity dispersion
profile of the lensing galaxy. In order to predict this value based on a set of galaxy parameters, 
we start with an expression, derived from the spherical Jeans equation, for $\sigma^2(r)$, 
the radial velocity dispersion of the luminous matter (i.e., stars) \citep{Binney80}:
\begin{equation}
\sigma^2_r(r) =  \frac{G\int_r^\infty dr' \ \nu(r') M(r') (r')^{2 \beta - 2} }{r^{2\beta} \nu(r)}~~~,
\label{eq:binney}
\end{equation}
for the case where the velocity anisotropy parameter, $\beta$, is a constant. 
Note that the use of Equation (\ref{eq:binney}) is based on the assumption that the 
relationship between stellar number density and stellar luminosity density is spatially constant, 
an assumption unlikely to be violated appreciably within the effective radius of the 
early-type lens galaxies under consideration. Using the mass 
density profile in Equation (\ref{eq:rhopl}), it is straightforward to show that the relation between 
the mass contained within a spherical radius $r$ and $M_E$ is
\begin{equation}
M(r) = \frac{2}{\sqrt{\pi} \lambda(\alpha)} \left(\frac{r}{R_E}\right)^{3 - \alpha} M_E ~~~,
\end{equation}
where we have defined the ratio of gamma functions
\begin{equation}
\lambda(x) = \Gamma \left(\tfrac{x-1}{2}\right) / \Gamma \left(\tfrac{x}{2}\right) ~~~.
\label{eq:lambda}
\end{equation}
Therefore, after evaluating the integral in Equation (\ref{eq:binney}) we find
\begin{equation}
\sigma^2_r(r) = \left[\frac{G M_E}{R_E} \right] \frac{2}{\sqrt{\pi}\left(\xi- 2 \beta \right) \lambda(\alpha)} 
\left(\frac{r}{R_E}\right)^{2 - \alpha}
\end{equation}
where, following \citet{Koopmans06}, we have defined $\xi = \delta + \alpha - 2$ as a convenient 
combination of the power-law exponents. 

Because all of the elliptical galaxy lenses in this study have been directly imaged with \textsl{HST}, 
we can measure their projected two-dimensional luminosity profiles.  Following \citet{BRB06},
we directly fit point-spread function (PSF) convolved two-dimensional power-law ellipsoid images to \textsl{HST} 
F814W imaging data over a circle of radius 1\farcs8 centered on the lens galaxies, adding 1 
to the best-fit two-dimensional power-law index so as to account for a deprojection into three 
dimensions to obtain our adopted $\delta$ values for each system. 

The velocity dispersions are measured from SDSS spectra. Each SDSS spectroscopic fiber subtends a
circle of radius 1\farcs5 on the sky, with atmospheric blurring adding an additional 
consideration.  The actual aperture weighting function $w(R)$ should thus be the convolution of the 
atmospheric seeing, represented by a Gaussian,
\begin{equation}
s(R) = \exp\left(-\frac{R^2}{2 \sigma_{\rm atm}^2}\right) ~~~,
\end{equation}
and the fiber aperture
\begin{equation}
a(R) = \Pi\left(\frac{R}{2R_{\rm fib}}\right) ~~~,
\end{equation}
where $\Pi$ is the rectangle function as defined in \citet{Bracewell}.
Then $w(R) = a(R) * s(R)$ which has the form
\begin{equation}
w(R) \propto e^{-R^2/2 \sigma_{\rm atm}^2} \int_0^{R_{\rm fib} / \sigma_{\rm atm}} d\eta \ \eta \ e^{-
\eta^2/2} \,I_0\left(\eta \frac{R}{\sigma_{\rm atm}}\right) ~~~,
\label{eq:conv}
\end{equation}
where $I_0$ is a modified Bessel function of the first kind. We have evaluated Equation (\ref
{eq:conv}) numerically, and found that as long as $R_{\rm fib} / \sigma_{\rm atm} \lesssim 1.5$, the 
convolution of the Gaussian and the top-hat will remain approximately Gaussian, where
\begin{equation}
\tilde{\sigma}_{\rm atm} \approx \sigma_{\rm atm} \sqrt{1 + \chi^2 / 4 + \chi^4 / 40}
\end{equation}
and $\chi = R_{\rm fib} / \sigma_{\rm atm}$. Thus, we will later use 
\begin{equation}
w(R) \approx  e^{-R^2/2 \tilde{\sigma}_{\rm atm}^2} ~~~,
\label{eq:wofr}
\end{equation}
for the aperture weighting function in Equation (\ref{eq:plsig}). 
For each lens in our analysis, we take for $\sigma_{\rm atm}$ the median value 
recorded by the spectroscopic guide cameras during the SDSS observations.

The actual velocity dispersion measured by the observations has been effectively 
luminosity-weighted along the line of sight and over the effective spectrometer aperture.
This averaging can be expressed mathematically as:
\begin{equation}
\left \langle \sigma^2_{*,\parallel} \right \rangle= \frac{\int_0^{\infty} dR \ R \ w(R) \int_{-\infty}^{\infty} 
d\mathcal{Z} \  \nu(r) \ \left(1 - \beta \tfrac{R^2}{r^2}\right)\sigma^2_r(r)}{\int_0^{\infty} dR \ R \ w(R) 
\int_{-\infty}^{\infty} d\mathcal{Z} \ \nu(r)}
\label{eq:sobs}
\end{equation}
where $w(R)$ is the observational aperture weighting function defined above in Equation (\ref{eq:wofr}). 
In Equation (\ref{eq:sobs}) the factor $\left(1 - \beta \tfrac{R^2}{r^2}\right)$ takes into account how 
the radial and tangential components of the velocity dispersion tensor project along the 
line of sight. This integral expression for $\left \langle \sigma^2_{*,\parallel} \right \rangle$
is analytic and, given the above definitions, it becomes
\begin{eqnarray}
\nonumber
\left \langle \sigma^2_{*,\parallel} \right \rangle &=& 
\left[\frac{2}{(1+\gamma)}\frac{c^2}{4} \frac{D_S}{D_{LS}} \theta_E \right] \frac{2}{\sqrt{\pi}} 
\frac{(2 \tilde{\sigma}_{\rm atm}^2)^{1 - \alpha/2}}{ (\xi - 2\beta)} \\ 
&&\times\left[\frac{\lambda(\xi) - \beta \lambda(\xi+2)}
{\lambda(\alpha)\lambda(\delta)}\right] \frac{ \Gamma(\tfrac{3-\xi}{2}) }{\Gamma(\tfrac{3 - \delta}{2}) }
~~~,
\label{eq:plsig}
\end{eqnarray}
where, again, $\lambda(x)$ is the ratio of gamma functions defined in Equation (\ref{eq:lambda}).
Henceforth, we abbreviate $\left \langle \sigma^2_{*,\parallel} \right \rangle$ as simply $\bar
{\sigma}_*^2$. One can immediately observe from Equation (\ref{eq:plsig}) that there are 
degeneracies among $\alpha$, $\beta$ and $\delta$.  The dependence upon 
$\Omega_{\Lambda}$ enters this relation by way of the ratio $D_{S} / D_{LS}$. 

\section{Determination of $\gamma$ and $\Omega_\Lambda$}
\label{sec:constraints}

As mentioned previously, the general approach to generating a constraint is to compare 
the velocity dispersion from the SDSS observations, $\sigma_{\rm SDSS}$, with  
the velocity dispersion calculated from a galaxy model $\bar{\sigma}_*$. By virtue of the
analysis in the previous section, we have
\begin{equation}
\bar{\sigma}_* = \bar{\sigma}_* (\alpha, \beta, \delta, \theta_E; \gamma, \Omega_\Lambda) ~~,
\label{eq:sigma_pars}
\end{equation}
where the semicolon separates the galaxy parameters and observables
($\alpha$, $\beta$, $\delta$, $\theta_E$)
for which we have measured or adopted values from the global parameters ($\gamma$, $
\Omega_\Lambda$) which we are seeking to constrain.

We will generate the constraint by considering  $\sigma_{\rm SDSS}$ 
and its corresponding uncertainty $\varepsilon_{\rm SDSS}$ to be a measurement
with Gaussian errors such that the probability density for
the observed value of $\sigma_{\rm SDSS}$ given a ``true''
value of $\bar{\sigma}_*$ is
\begin{equation}
P\left(\sigma_{\rm SDSS} | \bar{\sigma}_*\right) = \frac{1}{\sqrt{2 \pi} \varepsilon_{\rm SDSS}} \exp
\left[ -\frac{(\bar{\sigma}_* - \sigma_{\rm SDSS})^2}{2 \varepsilon_{\rm SDSS}^2}\right] ~~~.
\label{eq:alphabeta}
\end{equation}
Statistical errors on $\delta, $ $\theta_E$,
and the source and lens redshifts are negligible in comparison to
the velocity dispersion errors and the intrinsic variation of the galaxy parameters
$\alpha$ and $\beta$, and therefore we do not treat them explicitly.

We determine observational constraints for both $\gamma$ and $\Omega_\Lambda$ in a similar 
manner and thus the following discussion concerns our method of constraining a single parameter 
$X \in \{\gamma, \Omega_\Lambda\}$. In order to constrain $\gamma$ we assume a 
standard flat cosmology with $\Omega_\Lambda = 0.7$. In order to constrain $\Omega_\Lambda$, 
we assume that GR is correct and thus $\gamma = 1$.

We are interested in calculating the probability of our observations having yielded a certain value 
of galaxy velocity dispersion, 
$\sigma_{\rm SDSS}$, given a particular assumed value of the parameter $X$.
We incorporate the dependence upon $\delta$ and $\theta_E$ by
using the system-specific values determined from the
\textsl{HST} imaging as discussed 
in Sections \ref{sec:slacs} and \ref{sec:formalism}, although in what follows
we suppress $\delta$ and $\theta_E$ for notational convenience.
Because we cannot independently measure $\alpha$ and $\beta$ for individual lensing systems,
we consider these values to be drawn from Gaussian distributions $P(\alpha)$ and $P(\beta)$
with known mean and intrinsic scatter.
For a given lens, the probability density of interest can be written as
\begin{equation}
P(\sigma_{\rm SDSS} | X) = \int d\alpha \, P(\alpha) \int d\beta \, P(\beta) \,
P[\sigma_{\rm SDSS} | \bar{\sigma}_*(\alpha, \beta; X)]~.
\label{eq:pp}
\end{equation}
Assuming a flat prior $P(X)$ for $X$ over a range of interest, a universal value for $X$ in all systems, and 
statistical independence of the measurements of each system, the posterior probability for $X$ 
given the data
is then proportional to the product over the individual probabilities:
\begin{eqnarray}
\nonumber
P(X | \left\{\sigma_{\mathrm{SDSS},i} \right\}) &\propto&
P(\left\{\sigma_{\mathrm{SDSS},i} \right\} | X) \, P(X) \\
 ~ &\propto& P(X) \prod_{i=1}^{53} P(\sigma_{\mathrm{SDSS},i} | X) ~~~.
\end{eqnarray}

\begin{figure}[t]
\centering
\includegraphics[width = 0.48\textwidth]{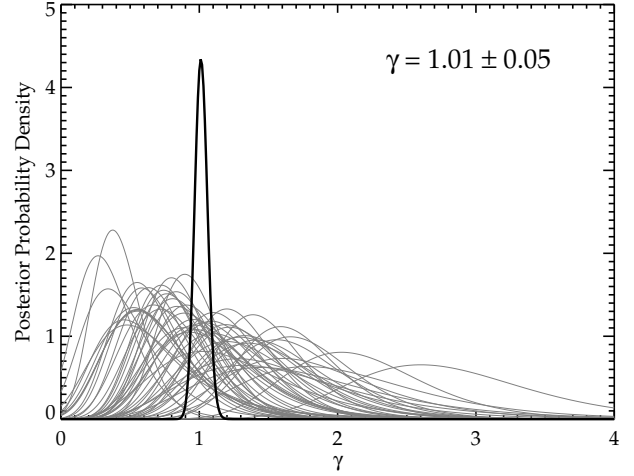}
\caption[Constraint on $\gamma$]{Constraint on $\gamma$ determined using the method 
discussed in the text. The gray curves represent the posterior PDF for $\gamma$ from each lens system. 
The black curve is the joint posterior PDF for all systems, the normalized product of the gray curves. 
(In this plot the joint PDF is scaled by a factor of one-half so as to relatively enhance the scale of the 
individual gray PDF curves.) A Gaussian fit to the joint posterior PDF gives $\gamma = 1.01 \pm 0.05$.  
}
\label{fig:gamma}
\end{figure}

We illustrate this analysis by selecting a fiducial set of parameters. For $\alpha$ and $\beta$ we 
choose a set of Gaussian distributions characterized by:
\begin{equation}
\begin{array}{lclclcl}
\avg{\alpha} &=& 2.00 & ; & \sigma_\alpha & = & 0.08 \\
\avg{\beta} & = & 0.18 & ; & \sigma_\beta & = & 0.13~~~.
\end{array}
\label{eq:alphabeta_pars}
\end{equation}
The cited 1-$\sigma$ ranges in $\alpha$ 
and $\beta$ are meant to indicate the intrinsic spreads in these quantities rather than 
uncertainties in the mean values.
The mean of $\alpha$ is fiducially taken to be simply 2.00, the slope of a singular isothermal sphere.
This is the same value as adopted in the analysis of \citet{Grillo08}.
This is also consistent with the value of $\avg{\alpha} = 1.96 \pm 0.08$ determined 
for the same sample of SLACS lenses by \citet{Koopmans09} via an ensemble aperture-mass
analysis, which is independent of any dynamical data
or modeling and hence does not introduce circularity into our logic
in proceeding to constraints on $\gamma$ and $\Omega_{\Lambda}$. The scatter in $\alpha$ and the 
distribution for $\beta$ are the same as used in \citet{BRB06}, and are taken to describe the 
distribution of mass-dynamical properties of the well-studied sample of nearby elliptical galaxies
from \citet{Gerhard01}.  \citet{Koopmans09} present a significantly higher value
for the intrinsic scatter in $\alpha$, but argue that this value should be regarded
as an upper limit given the likelihood of unmodeled systematic effects.
For comparison with these adopted values of $\avg{\alpha}$ and $\avg{\beta}$, 
our measured values of $\delta$ have a mean of $\avg{\delta} = 2.40$ and a standard deviation 
$\sigma_\delta = 0.11$.

Performing the analysis discussed above for $\gamma$, we find the resulting 
posterior probability density shown in Figure \ref{fig:gamma}. 
A fit to a Gaussian gives $\gamma = 1.01 \pm 0.05$ (1 $\sigma$ confidence). The 
result is consistent with $\gamma = 1$ and GR. Compared to the previous \cite
{BRB06} result, $\gamma = 0.98 \pm 0.07$, the statistical errors are reduced by a factor of 
$\sim 1.5$ as a result of the increase in the number of lensing systems considered (15 vs.~53). 
However, as we discuss in the next section, the systematic errors may be comparable to, 
or even dominate over, these statistical uncertainties, 
thereby preventing a large improvement in the precision of $\gamma$ 
simply by increasing the sample size of gravitational lens systems.

\begin{figure}
\centering
\includegraphics[width = 0.45 \textwidth]{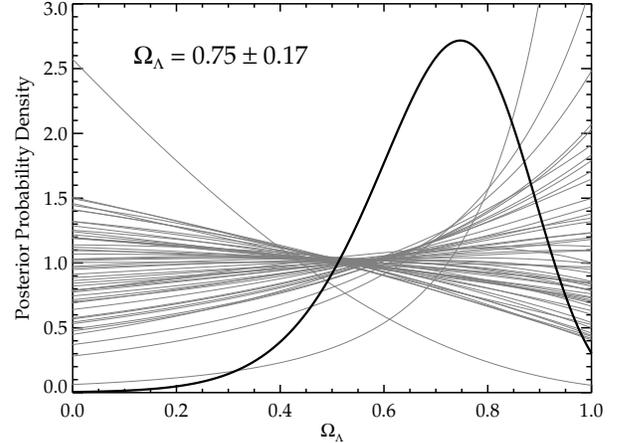}
\caption[Constraint on $\Omega_\Lambda$]{Constraint on $\Omega_\Lambda$ determined as 
discussed in the text. The gray curves represent the individual posterior PDFs of $\Omega_
\Lambda$ from each lens system. The black curve is the joint posterior PDF, the normalized product of 
the gray curves. A Gaussian fit to the joint PDF gives $\Omega_\Lambda = 0.75 \pm 0.17$.}
\label{fig:devol}
\end{figure}

We now set $\gamma =1$ and establish a constraint on $\Omega_\Lambda$. Because the 
calculation, and in particular the ratio $D_{\rm S}/D_{\rm LS}$ contained in Equation (\ref
{eq:plsig}), is not very sensitive (at more than the $\sim 10\%$ level) to cosmological parameters, 
we restrict our attention to flat universes with $\Omega_M + \Omega_\Lambda = 1$.  The resulting 
posterior probability density for $\Omega_\Lambda$ is shown in Figure \ref{fig:devol}. The peak 
occurs at $\Omega_\Lambda = 0.75$ and the distribution has a best-fit Gaussian width (1\,$\sigma
$) of $0.17$.  $\Omega_\Lambda = 0$ is clearly excluded.
This constraint is consistent with the results from the cosmic microwave background, galaxy 
clusters, and type Ia supernovae. The recent WMAP5 data in combination with type Ia supernovae 
and baryon acoustic oscillations indicate $\Omega_\Lambda = 0.726 \pm 0.015$ and is also 
consistent with our assumption of a flat universe, giving the constraint $-0.0179 < \Omega_k < 
0.0081$ \citep{WMAP5}.

\begin{figure}[h]
\centering
\includegraphics[width = 0.45\textwidth]{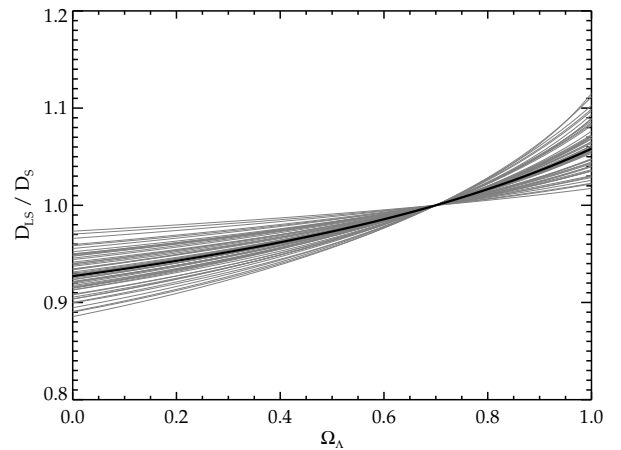}
\caption{Sensitivity of the ratio $D_{\rm LS}/D_{\rm S}$ to the cosmological parameter $\Omega_
\Lambda$. The gray curves shown are each of the 53 SLACS lenses. The median is shown by the 
thick black curve. All curves are normalized to a value of unity at a nominal value of $\Omega_
\Lambda = 0.7$.}
\label{fig:sensitivity}
\end{figure}

One can see from Figure \ref{fig:devol} that none of the individual posterior 
probability density function (PDF) curves peak: all 
are either strictly increasing or decreasing. This bimodality occurs because of the relative 
insensitivity of $D_{\rm S}/D_{\rm LS}$ to $\Omega_\Lambda$. A quick test shows that a reduction 
of the errors on $\sigma_{\rm SDSS}$ by a factor of $\sim 2$ would induce a few of the flatter 
likelihood curves to peak. However, for most systems, there is no value of $\Omega_\Lambda \in 
[0,1]$ such that $\bar{\sigma}_*(\avg{\alpha}, \avg{\beta}, \delta; \Omega_\Lambda) = \sigma_{\rm 
SDSS}$, which leads to the monotonic likelihood curves.

One may also observe that the posterior PDF curves for two of the systems rise or fall more sharply 
than the others. These two lens systems (J0252+0039 and J0737+3216) are among the high end of 
the SLACS sample in terms of lens redshift ($z_L \sim 0.3$) but appear to be otherwise unremarkable.
Neither lens strongly influences the final result: removing one or the other system shifts the peak 
posterior probability for $\Omega_\Lambda$ by about three-quarters of a standard deviation, and 
removing both leaves the peak of the posterior probability approximately unchanged.

To illustrate the dependence of the ratio $D_{\rm LS}/D_{\rm S}$ to the cosmological parameter $
\Omega_\Lambda$, we show in Figure \ref{fig:sensitivity} the ratio as a function of the assumed value 
of $\Omega_\Lambda$ for the 53 SLACS lenses. The curves are all normalized so that they yield a 
ratio of unity for a nominal value of $\Omega_\Lambda = 0.7$.  We see clearly from the curves in 
Figure \ref{fig:sensitivity} that the ratio of distances typically varies by $\pm 7\%$ and rarely by more 
than $\pm 10\%$. The fractional statistical error in the inferred distance ratios for each system will 
be roughly 10--15\%, or twice the fractional error in the measured velocity dispersion 
($D_S / D_{LS} \propto  \bar{\sigma}^2_*$ ; see Equation (\ref{eq:plsig})). Therefore 
a determination of $\Omega_\Lambda$ from an individual lens system can only be made at a 
``less-than-one-sigma'' level.  As previously noted by \citet{Grillo08},
many lenses are required to obtain a statistically significant 
determination of $\Omega_\Lambda$.

\section{The Effect of Systematics}
\label{sec:systematics}

\begin{figure}[t]
\centering
\includegraphics[width = 0.48\textwidth]{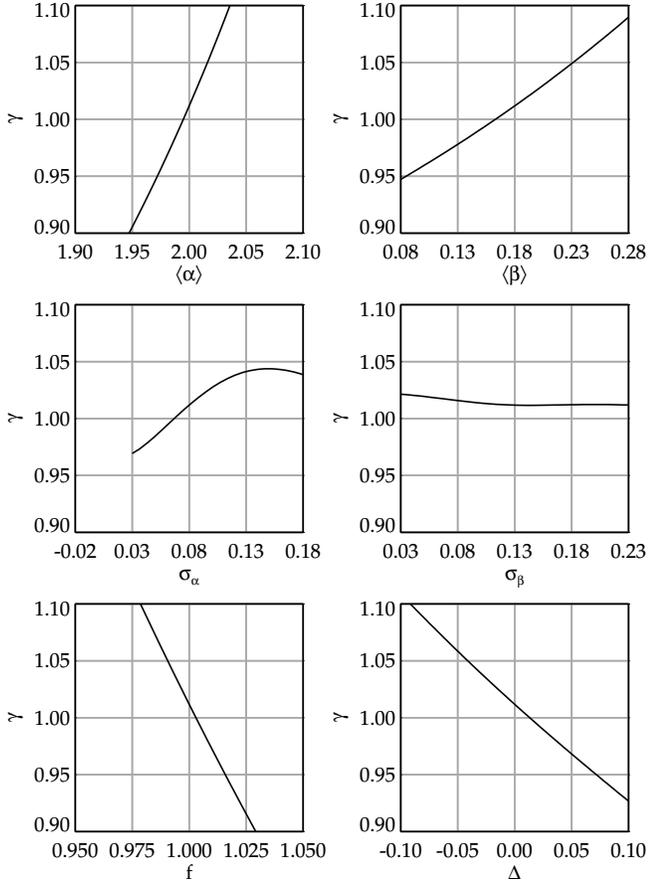}
\caption[]{Derived value of the post-Newtonian parameter, $\gamma$, as a function of each of 
the six lens parameters, as defined in the text, centered on their fiducial value. As one parameter 
is varied, the others are held at their fiducial values.}
\label{fig:gammafp}
\end{figure}

Figure \ref{fig:gamma} shows the posterior PDF for the post-Newtonian
parameter $\gamma$ using fiducial 
values for the lens model parameters: $\avg{\alpha} = 2.00$, $\sigma_\alpha = 0.08$, 
$\avg{\beta} = 0.18$ and $\sigma_\beta = 0.13$ (see Equation (\ref{eq:alphabeta_pars})), 
the measured slopes of the light profiles, $\delta$, and 
the values of the measured velocity dispersions, $\sigma_{SDSS}$.  
The figure provides a clear estimate of the statistical uncertainty in $\gamma$, 
such that $\gamma = 1.01 \pm 0.05$ ($1 \, \sigma$ confidence). 
Similarly, Figure \ref{fig:devol} shows the posterior PDF for the $\Omega_\Lambda$ 
parameter using the same fiducial values for the lens model parameters as described here for 
determining $\gamma$, which yields a value for $\Omega_\Lambda$ and its statistical uncertainty,
$\Omega_\Lambda = 0.75 \pm 0.17$.

However, each of the six lens parameters discussed above is susceptible to systematic error 
in either the adopted mean values ($\langle \alpha \rangle$ and $\langle \beta \rangle$), 
the assumed intrinsic widths ($\sigma_\alpha$ and $\sigma_\beta$), 
or in the directly measured parameters ($\delta$ and $\bar{\sigma}_{\rm SDSS}$).
It is therefore of crucial importance to test the sensitivity of our results for $\gamma$ and $\Omega_
\Lambda$ to systematic shifts in the parameters associated with the lensing galaxies and to judge 
whether these sensitivities pose immediate and fundamental limitations to the precision of our 
constraints.  To this end, we map out the dependence of the peak value of $\gamma$ and 
$\Omega_\Lambda$ (e.g., from results that are analogous to those shown in Figures
\ref{fig:gamma} and \ref{fig:devol}) on the lens parameters.  In particular, we vary $\langle \alpha 
\rangle$ over the range $1.90-2.10$ and $\langle \beta \rangle$ over the range $0.08 - 0.28$. 
We explore the effect of the assumed intrinsic widths by varying $\sigma_\alpha$ 
over the range $0.03 - 0.18$ and $\sigma_\beta$ over the range $0.03 - 0.23$.
Next we shift all the measured values of dispersion velocity, $\sigma_{\rm SDSS}$ by a
multiplicative factor, $f$, taken between 0.95 and 1.05, 
as well as all the measured slopes of the light distribution, $\delta$, by an additive
amount $\Delta$ ranging from $-0.10$ to 0.10.  As we vary each of the these parameters in turn, 
the other parameters are held fixed at their fiducial values.

\begin{figure}
\centering
\includegraphics[width = 0.48\textwidth]{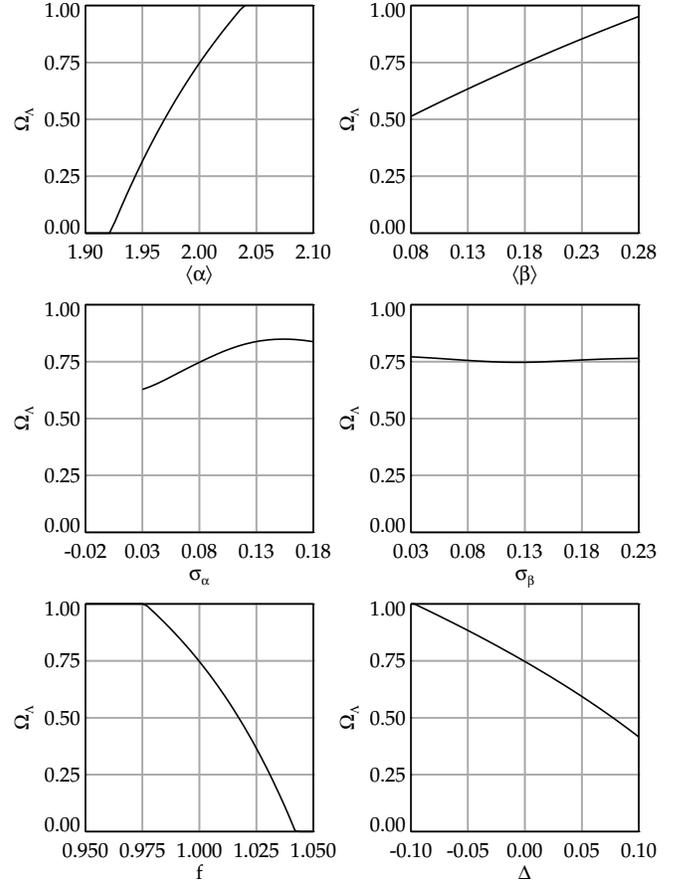}
\caption[]{Derived value of $\Omega_\Lambda$ as a function of each of the six lens 
parameters, as defined in the text, centered on their fiducial value. As one parameter is varied, the 
others are held at their fiducial values.}
\label{fig:omegafp}
\end{figure}

The results for the sensitivity of $\gamma$ to systematic variations 
in the six lens parameters are shown in Figure \ref{fig:gammafp}.  In each of the six panels the 
center of the plot is located at $\gamma = 1$ and at the fiducial value of the lens parameter
being varied: $\langle \alpha \rangle = 2.00$, $\langle \beta \rangle = 0.18$, 
$\sigma_\alpha = 0.08$, $\sigma_\beta = 0.13$, $f = 1.0$, and $\Delta = 0.0$, respectively.
The values of the slopes of these curves (at the 
fiducial point) are summarized in Table~\ref{tbl:derivs}.
The analogously determined sensitivity of $\Omega_\Lambda$ to systematic variations in the six lens 
parameters is shown in Figure \ref{fig:omegafp},
with the corresponding slopes also given in Table~\ref{tbl:derivs}.  The results 
are closely related to those for $\gamma$, except that they are typically a factor of $\sim$3 times 
larger in the case of $\Omega_\Lambda$ than for $\gamma$.  This is due to the fact that matching 
the model $\langle \sigma_{*,||} \rangle$ (as defined in Equation \ref{eq:plsig}) to the measured $\sigma$
is much more sensitively dependent on $\gamma$ than it is on $\Omega_\Lambda$.  Roughly
speaking, $\Omega_\Lambda$ is about three times more difficult to 
determine precisely than $\gamma$ using strong lensing.

In general, one can see from Figures \ref{fig:gammafp} and \ref{fig:omegafp}
and Table~\ref{tbl:derivs} that constraints on $\gamma$ and $\Omega_{\Lambda}$ are
quite sensitive to small systematic shifts in the adopted lens-galaxy parameters.
For example, a shift in the mean value of the slope 
of the mass profile of 0.04 in $\langle \alpha \rangle$
leads to a shift in $\gamma$ of $0.04 \times 2.3 \simeq 0.09$,
a change that is nearly twice the statistical uncertainty associated
with our $\gamma$ measurement.
Such a departure of $\langle \alpha \rangle$ from the value of 2.00,
which amounts to just one-half of the
adopted intrinsic scatter, is entirely
plausible given the variation in this quantity as deduced from independent
analyses and data sets (e.g., \citealt{Gerhard01, Rusin03, Rusin05, SLACSVII, Koopmans09}).
More crucially, this same shift in $\langle \alpha \rangle$ would produce
a change in $\Omega_{\Lambda}$ of 0.29, a nearly fatal degradation of our
constraint on this cosmological parameter.  Shifts of comparable magnitude
in $\gamma$ and $\Omega_{\Lambda}$
would be caused by a systematic error of 2.5\% in the measured velocity dispersions,
(i.e., $f = 1 \pm 0.025$) which is well within the range of uncertainty due to possible 
mismatch between the stellar populations of the lens galaxies and the
stellar spectrum template set used for the measurement of $\sigma_{\mathrm{SDSS}}$.

Though the effects are a bit less dramatic,
similar statements hold for the sensitivity of the
$\gamma$ and $\Omega_{\Lambda}$ results to the adopted
intrinsic mass-slope scatter $\sigma_\alpha$, the mean
velocity anisotropy $\langle \beta \rangle$, and the
systematic uncertainty in the light-profile slope $\Delta$.
With respect to the last quantity, we note that although
the power-law luminosity profile model that we have adopted
could be improved upon with a different choice such as
deVaucouleurs or S\'{e}rsic, the best-fit models from
these classes still show significant departures from the
\textsl{HST} imaging data.
Thus the sensitivity of $\gamma$ and $\Omega_{\Lambda}$ constraints
to systematic errors in the adopted luminosity profile is unlikely
to improve much beyond the level presented here.  Even massively
parametric models for the luminosity profile such as the
radial B-spline \citep{SLACSI, SLACSV} would be afflicted
by systematic uncertainties related to ellipticity and
deprojection.  In addition, other currently unmodeled systematic
effects such as selection effects \citep{Dobler08, Mandelbaum08}
and environmental overdensities
\citep{Auger08, SLACSVIII, Guimaraes09},
which appear to contribute only minorly to analyses of the structure
of SLACS lenses under
the assumption of $\gamma=1$ and $\Lambda$CDM, could
have significant implications for the type of
constraints considered in the current work.

\begin{deluxetable}{cccc}

\tablecolumns{4}
\tablewidth{0.45 \textwidth}

\tablecaption{Derivatives with Respect to Lens Parameters}

\tablehead{\colhead{$Y$ \tablenotemark{a}} & 
		  \colhead{$Y_0$ \tablenotemark{b}} & 
	           \colhead{$d\gamma / dY$} & 
	           \colhead{$d\Omega_\Lambda / dY$}}

\startdata
$\avg{\alpha}$  & 2.00 &  2.3 & 7.3 \\
$\avg{\beta}$ & 0.18 & 0.64 & 2.3 \\
$\sigma_\alpha$ & 0.08 & 0.85 & 2.4 \\
$\sigma_\beta$ & 0.13 & -0.03 & 0.06\\
$f$ & 1.00 & -4.0 & -12 \\
$\Delta$ & 0.00 & -0.9 & -2.9
\enddata

\tablenotetext{a}{All lens parameters are dimensionless and are defined in the text.}
\tablenotetext{b}{The fiducial value at which the derivatives are evaluated.}

\label{tbl:derivs}
\end{deluxetable}

From this analysis, we conclude that the systematics of the lens parameters must be controlled to a 
much tighter degree than currently appears possible before a more 
precise value of the post-Newtonian parameter $\gamma$ can be determined, and before a 
meaningful independent determination of $\Omega_\Lambda$ can be derived from strong 
gravitational lensing.

\section{Summary and Conclusions}
\label{sec:summcon}

In this work we have shown that, for a well-motivated set of fiducial lens-galaxy parameters, the 
current sample of 53 SLACS gravitational lensing galaxies can constrain either the post-Newtonian 
parameter $\gamma$ (on kpc length scales), or the cosmological parameter $\Omega_\Lambda$.  
The statistical errors are such that the measurements provide an interesting level of precision ($
\sim$0.05 in $\gamma$ and $\sim$0.17 in $\Omega_\Lambda$). We have paid particular attention 
to quantifying the effects of systematic uncertainties in the lens parameters on the determination of 
$\gamma$ and $\Omega_\Lambda$. These systematic errors are likely to be of equal or greater 
magnitude than the statistical uncertainties, so that the final result is largely limited by
our imperfect knowledge of
the galaxy parameters used in the analysis.  In particular, the derivatives of both $\gamma$ and $
\Omega_\Lambda$ with respect to several important lens parameters (enumerated in \S 5) have been 
calculated; these are displayed graphically in Figures \ref{fig:gammafp} and \ref{fig:omegafp} and 
tabulated in Table \ref{tbl:derivs}.

The larger lens sample we use as compared to that of \citet{BRB06} has enabled us to reduce the 
statistical error on $\gamma$ by a factor of $\sim$1.5 (from 0.07 to 0.05).  However, since
the uncertainty in $\gamma$ due to the systematic errors in the lens parameters is likely to be 
at least as large as the statistical errors, simply increasing the lens sample will not improve 
the constraints on $\gamma$ until the lens parameters are better understood.

Measurements of $\Omega_\Lambda$ based on strong lensing involve different sets of 
assumptions than other experimental methods, and therefore could provide a useful independent estimate of 
this important cosmological parameter.  However, our results show that in order for strong lensing 
by galaxies to become effective in determining $\Omega_\Lambda$, our independent knowledge 
of lens-galaxy mass-dynamical structure will have to improve by an order of magnitude.

Future observations and analysis have some prospect of improving this situation.
High-resolution ground-based spectroscopy with large telescopes and good seeing can reduce 
both the statistical and systematic errors on the measured velocity dispersions (\citealt{Czoske08, 
Barnabe09}), and may indeed be able to provide a measurement of the average velocity 
anisotropy for some systems.  A more detailed, spatially resolved dynamical analysis of nearby 
galaxies such as from the SAURON survey (\citealt{SAURONIII, SAURONIV}) could be used to 
derive more accurate prior distributions on lens-galaxy mass-model parameters, providing that the 
differing selection effects between SLACS and SAURON samples can be effectively controlled.  
Finally, system-by-system constraints on galaxy mass profiles
using higher-order information contained in lensing 
data alone (e.g., \citealt{Dye05, Willis06, Dye08}) could be used to remove some of the need for 
independent priors.

\acknowledgments

Support for \textsl{HST} programs No.\ 10174, No.\ 10494,
No.\ 10587, No.\ 10798, and No.\ 10886 was provided by NASA through grants from
the Space Telescope Science Institute,
which is operated by the Association of Universities for
Research in Astronomy, Inc., under NASA contract NAS 5-26555.

Funding for the SDSS and SDSS-II was provided by the Alfred P. Sloan Foundation,
the Participating Institutions, the National Science Foundation,
the U.S. Department of Energy, the National Aeronautics and Space Administration,
the Japanese Monbukagakusho, the Max Planck Society,
and the Higher Education Funding Council for England.
The SDSS was managed by the Astrophysical Research
Consortium for the Participating Institutions.

\bibliography{gamma_Omega}

\end{document}